\documentclass{elsart}
\usepackage{amsfonts,amsmath,amssymb,array,graphicx,tabularx}
\usepackage[mathscr]{eucal}

\newcommand{\diag}{\mathop{\rm diag}\nolimits}
\newcommand{\ch}{\mathop{\rm cosh}\nolimits}

\begin{document}

\begin{frontmatter}



\title{Amplitude equations for a system with thermohaline convection\thanksref{GOV}}


\author{S. B. Kozitskiy}\ead{skozi@poi.dvo.ru}

\address{Il'ichev Pacific oceanological institute, Baltiyskaya St. 43, Vladivostok, 41, 690041, Russia }
\thanks[GOV]{This is a preliminary and modified variant of the paper, published in Journal of Applied Mechanics and
Technical Physics, Vol. 41, No. 3, 2000, P.~429--438.}

\journal{JAMTP}\volume{41}\issue{3}

\setcounter{firstpage}{429} \setcounter{lastpage}{438}
\pubyear{2000}

\begin{abstract}
The multiple scale expansion method is used to derive amplitude
equations for a system with thermohaline convection in the
neighborhood of Hopf and Taylor bifurcation points and at the double
zero point of the dispersion relation. A complex Ginzburg-Landau
equation, a Newell-Whitehead-type equation, and an equation of the
$\varphi^4$ type, respectively, were obtained. Analytic expressions
for the coefficients of these equations and their various asymptotic
forms are presented. In the case of Hopf bifurcation for low and
high frequencies, the amplitude equation reduces to a perturbed
nonlinear Schr\"odinger equation. In the high-frequency limit,
structures of the type of ``dark" solitons are characteristic of the
examined physical system.
\end{abstract}

\end{frontmatter}



\section*{Introduction.}
In the 1980-1990s, a number of papers devoted to the formation of
structures in the neighborhood of Hopf bifurcation points for
systems with translational invariance along the horizontal appeared
in the literature on double-diffusive convection. Oscillations in
such systems can give rise to waves of various types (for example,
standing, traveling, modulated, and random), which are conveniently
studied by constructing amplitude equations \cite{Dean:1987}. An
amplitude equation for a system with convection was first obtained
by Newell and Whitehead \cite{Newell:1969}. It describes
two-dimensional thermal convection and has the form of a generalized
Ginzburg-Landau equation. Coulett et al. \cite{Coullet:1985}
proposed a system of Ginzburg-Landau equations that describes
traveling, double-diffusive waves propagating on both sides in a
liquid strip which is infinite along the horizontal:
\begin{eqnarray} \label{Cou}
\begin{array}{l}
(\partial_t+s\partial_x)A_R = (c_0+i c_1) A_R +
(c_2+i c_3)\partial^2_x A_R -  \\
\qquad \qquad - (c_4+i c_5)|A_R|^2 A_R - (c_6+i c_7)|A_L|^2 A_R,\\
(\partial_t-s\partial_x)A_L = (c_0+i c_1) A_L +
(c_2+i c_3)\partial^2_x A_L - \\
\qquad \qquad - (c_4+i c_5)|A_L|^2 A_L - (c_6+i c_7)|A_R|^2 A_L.
\end{array}
\end{eqnarray}
The form of these equations is postulated from general
considerations (such as considerations of symmetry); it is assumed
that the coefficients in these equations should be derived by
asymptotic methods from the partial equations describing a
particular physical system.

However, a thorough and well-founded derivation of amplitude
equations for double-diffusive systems is not available in the
literature there. In many papers, the form of the coefficients in
Eqs. (\ref{Cou}) is not discussed. In some papers, these
coefficients are obtained from various physical considerations.
Thus, Cross \cite{Cross:1988}, examining a system with convection
for binary mixtures in the limit of low Hopf frequencies, set the
coefficients $c_1, c_3, c_5$ and $c_7$ in Eqs. (\ref{Cou}) equal to
zero as a first, crude, approximation, motivating this by empirical
data indicating an analogy between the case considered and the case
of purely temperature convection. Clearly, such assumptions on the
form of the coefficients can be rigorously justified only in a
rigorous mathematical derivation of amplitude equations.

In papers on double diffusive convection of binary mixtures in bulk
and porous media, the Hopf frequency turns out to be unity in the
case of oscillatory convection. For thermohaline convection, it is
reasonable to consider the asymptotic behavior for the Hopf
frequency tending to infinity. In this limit, the amplitude equation
should become the nonlinear Schr\"odinger equations governing
internal waves in two-dimensional geometry.

In the present paper, using the derivative expansion method, which
is a version of the multiple-scale expansion method, we derive
amplitude equations for double-diffusive waves in two-dimensional,
horizontally infinite geometry in the neighborhood of the Hopf and
Taylor bifurcation points and the double zero point of the
dispersion relation. Idealized boundary conditions are used. In the
case of Hopf bifurcation, the amplitude equation for waves
propagating only in one direction is examined. Analytic expressions
for the coefficients of these equations are obtained. Their various
asymptotic forms are studied, and asymptotic forms of the amplitude
equations for various parameter values are discussed.


\section{Formulation of the Problem; Basic Equations.}
The initial equations describe two-dimensional thermohaline
convection in a liquid layer of thickness $h$ bounded by two
infinite, plane, horizontal boundaries. The liquid moves in a
vertical plane and the motion is described by the stream function
$\psi(t,x,z)$. The horizontal $x$ and vertical $z$ space variables
are used; the time is denoted by $t$. It is assumed that there are
no distributed sources of heat and salt, and on the upper and lower
boundaries of the regions, these quantities have constant values.
Hence, the main distribution of temperature and salinity is linear
along the vertical and does not depend on time. The variables
$\theta(t,x,z)$ and $\xi(t,x,z)$ describe variations in the
temperature and salinity about this main distribution. There are two
types of thermohaline convection: the finger type, in which the
warmer and more saline liquid is at the upper boundary of the
regions, and the diffusive type, in which the temperature and
salinity are greater at the lower boundary. In the present paper, we
study the second type.

The evolution equations in the Boussinesq approximation in
dimensionless form are a system of nonlinear equations in
first-order partial derivatives with respect to time that depend on
four parameters: the Prandtl number $\sigma$, the Lewis number
$\tau$ $(0<\tau<1)$, and the temperature ${\rm R}_T$ and salinity
${\rm R}_S$ Rayleigh numbers \cite{Huppert:1976a,Knobloch:1986}:
\begin{eqnarray} \label{maineq}
& & \left( \partial_t - \sigma \Delta\right)\Delta\psi + \sigma
 \left({\rm R}_S \partial_x \xi - {\rm R}_T \partial_x \theta
 \right) = - J( \psi,\Delta \psi),  \nonumber \\
& & \left(\partial_t - \Delta \right)\theta - \partial_x\psi =
 - J( \psi,\theta ), \\
& & \left(\partial_t - \tau \Delta \right) \xi - \partial_x
 \psi =  - J( \psi,\xi). \nonumber
\end{eqnarray}
Here we have introduced the Jacobian
$J(f,g)=\partial_{x}{f}\partial_{z}{g}-
\partial_{x}{g}\partial_{z}{f}$.
The boundary conditions for the dependent variables are chosen to be
zero, which implies that the temperature and salinity at the
boundaries of the region are constant, the vortex vanishes at the
boundaries, and the boundaries are impermeable:
\begin{eqnarray} \label{econ}
\psi = \partial_z^2 \psi = \theta = \xi = 0
  \mbox{ при } z=0,\,\, 1.
\end{eqnarray}
In the literature, these boundary conditions are usually called
free-slip conditions or simply free conditions since the horizontal
velocity component at the boundary does not vanish.

The space variables are made nondimensional with respect to the
thickness of the liquid layer $h$. As the time scale, we use the
quantity $t_0={h^{2}}/{\chi}$, where x is the thermal diffusivity of
the liquid. The vertical and horizontal components of the
liquid-velocity field are defined by the formulas
\begin{eqnarray}
v_{z} = \frac{\chi}{h}{\partial_{x}}{\psi},  \qquad
v_{x} = - \frac{\chi}{h}{\partial_{z}}{\psi}. \nonumber
\end{eqnarray}
The dimensional temperature $T$ and salinity $S$ are given by the
relations
\begin{eqnarray}
& & T(t,x,z)=T_{-}+\delta {T}\left[1-z+\theta(t,x,z)\right],
 \nonumber \\
& & S(t,x,z)=S_{-}+\delta {S}\left[1-z+\xi(t,x,z)\right].
 \nonumber
\end{eqnarray}
Here $\delta T=T_{+}-T_{-}$, $\delta S=S_{+}-S_{-}$, and $T_{+}$,
$T_{-}$ and $S_{+}$, $S_{-}$ are the temperatures and salinities on
the lower and upper boundaries of the region, respectively. The
temperature and salinity Rayleigh numbers can be expressed in terms
of the parameters of the problem:
\begin{eqnarray}
{\rm R}_T = \frac{{g}{\alpha}{h^{3}}}{\chi\nu}{\delta}{T},
\qquad {\rm R}_S = \frac{{g}{\gamma}{h^{3}}}{\chi\nu}{\delta}{S},
 \nonumber
\end{eqnarray}
Here $g$ is the acceleration of gravity, $\nu$ is the viscosity of
the liquid, and $\alpha$ and $\gamma$ are the temperature and
salinity coefficient of cubic expansions.


\section{Dispersion Relation and Its Consequences.}
We consider a system of partial differential equations that is
derived by linearization of the initial system ($\ref{maineq}$) in
the neighborhood of the trivial solution:
\begin{eqnarray} \label{eq1}
& & \left(\partial_t - \sigma\Delta\right)\Delta\psi + \sigma
 \left({\rm R}_S \partial_x \xi - {\rm R}_T \partial_x \theta
 \right) = 0, \nonumber \\
& & \left(\partial_t - \Delta \right)\theta - \partial_x\psi = 0,
 \\
& & \left(\partial_t - \tau \Delta \right) \xi - \partial_x
 \psi = 0. \nonumber
\end{eqnarray}
These equations are solved subject to boundary conditions
($\ref{econ}$) by the method of separation of variables. We seek a
solution in the form
\begin{equation} \label{sol}
\boldsymbol{\varphi}=\left[{\bf A}_1\exp(\lambda t-i\beta x)
+{\bf\bar A}_1\exp(\bar\lambda t+i\beta x)\right]\sin(\pi z).
\end{equation}
Here the bar denotes complex conjugation, $\boldsymbol{\varphi}$ is
the vector of the basic dependent quantities
$\boldsymbol{\varphi}=(\psi,\theta,\xi)$, $\beta$ is the horizontal
wavenumber, and ${\bf A}_1=(a_{A1}, a_{T1}, a_{S1})$ is the
amplitude vector. For $a_{A1}$ we use the notation $A\equiv a_{A1}$.

Substitution of ($\ref{sol}$) into ($\ref{eq1}$) gives a system of
algebraic equations for the variables $a_{A1}, a_{T1}$ and $a_{S1}$.
The condition for the existence of solutions of this system takes
the form of an algebraic equation of the third order in $\lambda$
\cite{Knobloch:1986}:
\begin{eqnarray} \label{disp}
\lambda^3+{\left( 1+\tau+\sigma \right)}{k^2}{\lambda^2}+
\left[\left(\tau+\sigma+\tau\sigma\right)+\sigma\left( r_{S}-r_{T}
\right)\right]{k^4}{\lambda}+ \nonumber \\
+ \sigma\left(r_{S}-\tau {r_{T}}+\tau\right){k^6}=0.
\end{eqnarray}
Here we introduced the wavenumber $k^2=\pi^2+\beta^2$, and the
normalized Rayleigh numbers $r_T={\rm R}_T/{\rm R}^{*}$ and
$r_S={\rm R}_S/{\rm R}^{*}$, where ${\rm
R}^{*}={k^4}{\left({k}/\beta \right)}^2$ is the Rayleigh number, for
which there is loss of stability of the steady state for purely
temperature convection.

Equation (\ref{disp}) has three roots, two of which can be complex
conjugate. In the physical system considered, loss of stability
occurs when with change in the bifurcation parameters $r_T$ and
$r_S$, one or several roots pass through zero or gain a positive
real part if they are complex.

\begin{figure}[htb]
\begin{center}
\begin{picture}(215,160)
\put(15,10){\vector(0,1){150}}
\put(15,10){\vector(1,0){200}}
\put(40,10){\line(1,1){50}}
\put(90,60){\line(1,2){50}}
\put(90,60){\circle*{3}}
\put(95,55){$C$}
\put(0,155){$r_S$}
\put(210,0){$r_T$}
\put(50,105){I}
\put(165,60){II}
\put(55,40){{\it 1}}
\put(100,105){{\it 2}}
\end{picture}
\end{center}
\caption{Plane of the parameters $r_T$ and $r_S$.}\label{r1}
\end{figure}
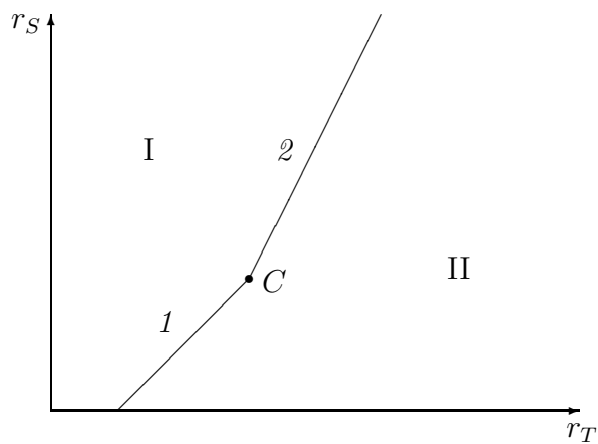

In the plane of the parameters $r_T$ and $r_S$ (see Fig.~\ref{r1}),
it is possible to distinguish regions I and II, on whose boundary
there is loss of stability. The boundary itself consists of two
rectilinear segments. On segment {\it 1}, Taylor bifurcation is
observed when one of the roots of the dispersion relation passes
through zero, which gives rise to steady drum-type convection. On
segment {\it 2}, Hopf bifurcation takes place when the real parts of
two complex conjugate roots become positive. As a result,
oscillatory convection occurs. The segments adjoin at the point $C$,
at which the dispersion relation (\ref{disp}) has a double root. At
this point, the parameter values are defined by
\begin{eqnarray}
r_{T1}=\frac{1}{\sigma}\frac{\tau+\sigma}{1-\tau}, \qquad
r_{S1}=\frac{\tau^2}{\sigma}\frac{1+\sigma}{1-\tau}. \nonumber
\end{eqnarray}
The straight lines on which Taylor and Hopf bifurcations are
observed, are given, respectively, by the equations
\begin{eqnarray}
r_T = \frac{1}{\tau}r_S + 1, \qquad
r_T = 1+\frac{\tau}{\sigma}{\left( 1+\tau+\sigma \right) }+
\frac{\tau+\sigma}{1+\sigma}{r_S}. \nonumber
\end{eqnarray}
The oscillation frequency of oscillatory convection is determined by
the imaginary part $\lambda$ and is expressed in terms of the
reduced frequency $\Omega$ as ${\rm Im}(\lambda)=\Omega k^2$, and
$\Omega$ is, in turn, calculated from the formula
\begin{eqnarray}
\Omega^2=-\tau^2+\frac{1-\tau}{1+\sigma}{\sigma}{r_{S}}, \qquad
\lambda = {i}\Omega {k^2}. \nonumber
\end{eqnarray}
Below, the reduced frequency $\Omega$ is called the Hopf frequency.


\section{Slow Variables and Expansion of the Solutions.}
We consider the equations of double-diffusive convection in the
neighborhood of a certain bifurcation point, for which the
temperature and salinity Rayleigh numbers are denoted by ${\rm
R}^{*}_T$ and ${\rm R}^{*}_S$, respectively. In this case, the
Rayleigh number is written as
\begin{eqnarray}
{\rm R}_T={\rm R}^{*}_T\left(1+{\varepsilon}^{2}\eta\right),
 \qquad
{\rm R}_S={\rm R}^{*}_S\left(1+{\varepsilon}^{2}\eta_S\right).
 \nonumber
\end{eqnarray}
The values of $\eta$ and $\eta_S$ are of the order of unity, and the
small parameter $\varepsilon$ shows how far from the bifurcation
point the examined system is located. To derive amplitude equations,
we use the derivative expansion method of
\cite{Dodd:1988,Nayfeh:1976}. We introduce the slow variables
\begin{eqnarray}
T_1 = \varepsilon t,\quad T_2=\varepsilon^2 t,\quad
X_1 = \varepsilon x. \nonumber
\end{eqnarray}
Next, into the basic equations ($\ref{maineq}$), we introduce the
extended derivative by the rules
\begin{eqnarray}  \label{deri}
\partial_t \rightarrow \partial_t + \varepsilon\partial_{T_1}
 + {\varepsilon}^{2}\partial_{T_2}, \qquad
\partial_x \rightarrow \partial_x + \varepsilon\partial_{X_1}.
\end{eqnarray}
The dependent variables are represented as series in the small
parameter:
\begin{eqnarray}
\boldsymbol{\varphi} = \sum_{n=1}^{3}\varepsilon^n
\boldsymbol{\varphi}_n(x,t,X_1,T_1,T_2) + O(\varepsilon^4).
 \nonumber
\end{eqnarray}
Substituting these expressions into Eqs. ($\ref{maineq}$) with
derivatives extended according to ($\ref{deri}$) and grouping terms
with the same power of $\varepsilon$, we obtain
\begin{eqnarray}
& & O(\varepsilon):\hphantom{^2}\quad
 L^{*}\boldsymbol{\varphi}_1 = 0, \nonumber \\
& & O(\varepsilon^2):\quad L^{*}\boldsymbol{\varphi}_2 =
 -(L_1\partial_{T_1}-L_2\partial_{X_1})
 \boldsymbol{\varphi}_1 - {\bf M}_1\boldsymbol{\varphi}_1,
 \nonumber \\
& & O(\varepsilon^3):\quad L^{*}\boldsymbol{\varphi}_3 =
 - (L_1\partial_{T_1}-L_2\partial_{X_1})
\boldsymbol{\varphi}_2 - \nonumber \\
& & \qquad - (L_1\partial_{T_2}+L_3\partial^2_{X_1}+
 L_4\partial_{X_1}\partial_{T_1}+L_5)\boldsymbol{\varphi}_1
 - {\bf M}_2(\boldsymbol{\varphi}_1,\boldsymbol{\varphi}_2).
 \nonumber
\end{eqnarray}
Here the operators $L_1, L_3$ and $L_4$ have diagonal form: $\diag
L_1=(\Delta, 1, 1)$, $\diag L_3 =
(\partial_t-2\sigma\Delta-4\sigma\partial^2_x, -1, -1)$, $\diag
L_4=(2\partial_x, 0, 0)$; the operators $L^{*}$ and $L_2$ can be
written as
\begin{eqnarray}
L^{*} &=& \left(
\begin{array}{ccc}
(\partial_t-\sigma\Delta)\Delta & -\sigma {\rm R}^{*}_T\partial_x
& \sigma {\rm R}^{*}_S\partial_x \\
- \partial_x & (\partial_t-\Delta) & 0 \\
- \partial_x & 0 & (\partial_t-\tau\Delta) \\
\end{array} \right), \nonumber \\
L_2 &=& \left(
\begin{array}{ccc}
-2(\partial_t-2\sigma\Delta)\partial_x & \sigma {\rm R}^{*}_T
& -\sigma {\rm R}^{*}_S \\
1 & 2\partial_x & 0 \\
1 & 0 & 2\tau\partial_x \\
\end{array} \right). \nonumber
\end{eqnarray}
In the operator $L_5$, only the upper row is different from zero:
$(0,$ $-\sigma {\rm R}^{*}_T\eta\partial_x,$ $\sigma {\rm
R}^{*}_S\eta_S\partial_x)$. The vectors ${\bf
M}_i=(M_{Ai},M_{Ti},M_{Si})$ with nonlinear terms have the
components
\begin{eqnarray}
& & M_{A1}=J(\psi_1,\Delta\psi_1), \quad M_{T1} =
 J(\psi_1,\theta_1), \quad M_{S1}=J(\psi_1,\xi_1), \nonumber \\
& & M_{A2}=J(\psi_2,\Delta\psi_1)+J(\psi_1,\Delta\psi_2)+
 J(\psi_1,2\partial_x\partial_{X_1}\psi_1)+ \nonumber \\
& & \qquad + \partial_z\Delta\psi_1\partial_{X_1}\psi_1 -
 \partial_z\psi_1 \partial_{X_1}\Delta\psi_1, \nonumber \\
& & M_{T2}=J(\psi_1,\theta_2)+J(\psi_2,\theta_1)+
 \partial_z\theta_1\partial_{X_1}\psi_1-
 \partial_z\psi_1\partial_{X_1}\theta_1, \nonumber \\
& & M_{S2}=J(\psi_1,\xi_2)+J(\psi_2,\xi_1)+
 \partial_z\xi_1\partial_{X_1}\psi_1-
 \partial_z\psi_1\partial_{X_1}\xi_1. \nonumber
\end{eqnarray}
The systems obtained can be written in general form:
\begin{eqnarray}
L^{*}\boldsymbol{\varphi}_i = {\bf Q}^{'}_i+{\bf P}_i. \nonumber
\end{eqnarray}
Here the functions ${\bf Q}^{'}_i$ consist of terms that are in
resonance with the left side of the equations. The condition of the
absence of secular terms in solutions of similar systems of
equations is known (see \cite{Dodd:1988,Nayfeh:1976}) to reduce to
the requirement that the functions ${\bf Q}^{'}_i$ and the solutions
of the conjugate homogeneous equation
$(L^{*})^{\star}{\boldsymbol{\varphi}_{i}^{\star}} = 0$ be
orthogonal. We now derive the relation to which the condition of the
absence of secular terms reduces in this case and which will be used
below to derive amplitude equations. Let us consider the
inhomogeneous system of algebraic equations that is obtained from
($\ref {eq1}$) by choosing the single-mode form (\ref{sol}) and
substituting functions ${\bf Q}_i$ $=(Q_{Ai}$, $Q_{Ti}$, $Q_{Si})$
such that ${\bf Q}^{'}_i={\bf Q}_i\exp(\lambda t-i\beta x)+ {\bf\bar
Q}_i\exp(\bar\lambda t+i\beta x)$ into the right side of the
homogeneous system:
\begin{eqnarray} \label{eq_al}
\begin{array}{l}
(\lambda+\sigma{k^2})(-k^2)a_{Ai}
+ \sigma {\rm R}^{*}_T i\beta a_{Ti}
- \sigma {\rm R}^{*}_S i\beta a_{Si} = Q_{Ai}, \\
(\lambda+{k^2}) a_{Ti} + i\beta a_{Ai} = Q_{Ti}, \\
(\lambda+\tau {k^2}) a_{Si} + i\beta a_{Ai} = Q_{Si}. \\
\end{array}
\end{eqnarray}
The solvability condition for this system of equations is formulated
as the requirement that the right side be orthogonal to the solution
of the conjugate homogeneous system \cite{Nayfeh:1980}
$\left( 1, - i\beta\sigma {\rm R}^{*}_T/(\lambda+k^2),
 i\beta\sigma {\rm R}^{*}_S/(\lambda+\tau k^2) \right)$
and reduces to the equation
\begin{eqnarray} \label{eqls}
(\lambda+{k^2}){k^6}\sigma r^{*}_S Q_{Si} -
(\lambda+\tau {k^2}){k^6}\sigma r^{*}_T Q_{Ti} - \nonumber \\
- (\lambda+{k^2})(\lambda+\tau {k^2}) i\beta Q_{Ai} = 0.
\end{eqnarray}
For $\lambda=0$, this relation is simplified:
\begin{eqnarray} \label{eqlc}
\frac{1}{\tau}{r^{*}_S}{Q_{Si}}-{r^{*}_T}{Q_{Ti}}
- \frac{i\beta}{\sigma k^4}{Q_{Ai}}=0.  \nonumber
\end{eqnarray}


\section{Derivation of Amplitude Equations.}
Let us assume that the solution of the equations for
$\boldsymbol{\varphi}_1$ has the form ($\ref{sol}$) and the
amplitude of this solution now depends on the slow variables:
$A=A(T_1,X_1,T_2)$. Substituting this solution into the equations
for $\boldsymbol{\varphi}_2$ we obtain a system of equations of the
form ($\ref {eq_al}$) in which the functions ${\bf Q}_2$ are written
as
\begin{eqnarray}
& & Q_{A2} = {k^2}\partial_{T_1}A + i\beta\sigma
 \left(\frac{{\rm R}^{*}_S}{\lambda+\tau k^2}
 - \frac{{\rm R}^{*}_T}{\lambda+k^2}+
 4{k^2} + \frac{2\lambda}\sigma \right)\partial_{X_1}A,
 \nonumber \\
& & Q_{T2}=\frac{i\beta}{\lambda+k^2}\partial_{T_1}A+
 \left( 1-\frac{2\beta^2}{\lambda+k^2} \right)
 \partial_{X_1}A, \nonumber \\
& & Q_{S2}=\frac{i\beta}{\lambda+\tau k^2}\partial_{T_1}A +
 \left( 1 - \frac{2\tau\beta^2}{\lambda+\tau k^2} \right)
 \partial_{X_1}A.  \nonumber
\end{eqnarray}
For this algebraic system to be solvable, it is necessary that
condition ($\ref {eqls}$) be satisfied. At different bifurcation
points, this condition is formulated as different equations. We
consider successively the equations obtained from the solvability
condition of the indicated system at the bifurcation points
characteristic of the physical system considered.

In the last relations, we substitute the value of $\lambda$ at the
Hopf bifurcation point $\lambda=i\Omega k^2$ and set $k^2/\beta^2=3$
and $\beta=\pi/\sqrt{2}$, which is valid for the oscillation mode
that is the first to lose stability \cite{Huppert:1976a}. In
addition, we take into account the relations
\begin{eqnarray} \label{rtrs}
r^{*}_{T}=\frac{1}{\sigma}\frac{\sigma+\tau}{1-\tau}
\left(\Omega^2 + 1 \right), \qquad
r^{*}_{S}=\frac{1}{\sigma}\frac{\sigma+1}{1-\tau}
\left(\Omega^2 + \tau^2 \right). \nonumber
\end{eqnarray}
Then, Eq. ($\ref {eqls}$) can be written as
\begin{eqnarray} \label{pareq}
\partial_{T_{1}}{A} + \sqrt{2}\pi\Omega\partial_{X_1}{A} = 0
 \nonumber
\end{eqnarray}
and solved in general form by introducing the new slow variable
$X=X_1-\sqrt{2}\pi\Omega T_1$. If we assume that the amplitude
$A(X,T_2)$ depends on $X_1$ and $T_1$ only via $X$, this equation
becomes an identity.

In the other cases, where we consider the system at the Taylor
bifurcation point or at the double zero point, the solvability
condition ($\ref {eqls}$) has the form
\begin{eqnarray} \label{teq}
\frac{1}{\tau}(1-\tau)\left(r^{*}_T
 - \frac{\sigma+\tau}{\sigma (1-\tau)}
 \right)\partial_{T_1}{A} + 2i\beta\left(\frac{k^2}{\beta^2}
 - 3\right)\partial_{X_1}{A}=0.
\end{eqnarray}
If in this equation, as above, $k^2/\beta^2=3$, i.e., the least
stable oscillation mode is considered, then $\partial_{T_1}{A}=0$
holds for the case of Taylor bifurcation. In the case of the double
zero point, Eq. ($\ref {teq}$) is satisfied identically.


\section{Amplitude Equation at the Hopf Bifurcation Point.}
We now write the solution for $\boldsymbol{\varphi}_2$ with the
wavenumber for which there is loss of stability of the steady state:
\begin{eqnarray}
& & \boldsymbol{\varphi}_2 = \left[{\bf A}_2
 \exp(i\Omega k^2 t-i\beta x)
 + {\bf\bar A}_2\exp(-i\Omega k^2 t+i\beta x)\right]
 \sin(\pi z)+ \nonumber \\
& & \qquad \qquad \qquad +{\bf B}_2\sin(2\pi z). \nonumber
\end{eqnarray}
Here ${\bf A}_2=(a_{A2},a_{T2},a_{S2})$ and ${\bf B}_2 =
(0,b_{T2},b_{S2})$ are vectors that depend on the slow variables.
The components of these vectors have the values
\begin{eqnarray}
& & b_{T2}= - \frac{1}{6\pi}\frac{|A|^2}{1+\Omega^2}, \qquad
    b_{S2}= - \frac{1}{6\pi}\frac{|A|^2}{\tau^2+\Omega^2},
    \nonumber \\
& & a_{T2}=\frac{2}{9\pi^2}\frac{1}{1+i\Omega}\left(\partial_X{A}
    - \frac{3i\pi}{\sqrt{2}}{a_{A2}} \right),  \nonumber \\
& & a_{S2}=\frac{2}{9\pi^2}\frac{1}{\tau+i\Omega}\left(
    \partial_X{A} - \frac{3i\pi}{\sqrt{2}}{a_{A2}} \right).
    \nonumber
\end{eqnarray}
Using the solutions given above, we formulate a system of equations
from which it is possible to find $\boldsymbol{\varphi}_3$. This
system of equations, as the system for $\boldsymbol{\varphi}_2$, has
the form ($\ref{eq_al}$). Then, we can write the functions ${\bf
Q}_3$ as follows, retaining in them only terms with $A(X,T_2)$:
\begin{eqnarray}
& & Q_{A3}=\frac{3}{2}\pi^2\left\{\partial_{T_2}A -
\frac{1}{3}(4i\Omega+7\sigma)\partial^2_X A + \right.\nonumber \\
& & \qquad \left. + \frac{3\pi^2}{2(1-\tau)}
[(\sigma+1)(\tau-i\Omega)\eta_S-(\sigma+\tau)(1+i\Omega)\eta]
A\right\}, \nonumber \\
& & Q_{T3}=\frac{i\sqrt{2}}{3\pi}\,\frac{1}{1+i\Omega}\left[
\partial_{T_2}A-\frac{1}{3}(2i\Omega+5)\partial^2_X A+
\frac{\pi^2}{4}\,\frac{1}{1-i\Omega}A|A|^2\right], \nonumber \\
& & Q_{S3}=\frac{i\sqrt{2}}{3\pi}\,\frac{1}{\tau+i\Omega}\left[
\partial_{T_2}A-\frac{1}{3}(2i\Omega+5\tau)\partial^2_X A+
\frac{\pi^2}{4}\,\frac{1}{\tau-i\Omega}A|A|^2\right]. \nonumber
\end{eqnarray}
Condition ($\ref{eqls}$) for system ($\ref{eq_al}$) has the form
\begin{eqnarray}
(\sigma+1)(\tau-i\Omega)Q_{S3}-(\sigma+\tau)(1-i\Omega)
Q_{T3} - (1-\tau)\frac{i\beta}{k^4}{Q_{A3}}=0. \nonumber
\end{eqnarray}
After transformations, we find that the amplitude $A(X,T_2)$ should
satisfy the complex Ginzburg-Landau equation
\begin{eqnarray} \label{CGLE}
\partial_{T_2} A = \alpha_1 A + \beta_1 A|A|^2 +
\gamma_1 \partial_{X}^2 A.
\end{eqnarray}
The coefficients in this equation are defined by the formulas
\begin{eqnarray}
& & \alpha_1=\frac{3i\pi^2 [\eta_S(\sigma+1)
(\Omega^2+\tau^2)(i\Omega+1)-\eta (\sigma+\tau)
(\Omega^2+1)(i\Omega+\tau)] }
{4\Omega[i\Omega+(1+\sigma+\tau)](1-\tau)},  \nonumber \\
& & \beta_{1}= - \frac{i\pi^2}{8\Omega}, \qquad
\gamma_{1} = i\Omega +
2\frac{(\sigma+\sigma\tau+\tau)\Omega-i\sigma\tau}
{\Omega [i\Omega+(1+\sigma+\tau)]}. \nonumber
\end{eqnarray}


\section{Equation in the Form of a Perturbed Nonlinear Schr\"odinger Equation.}
For further investigation, the equation obtained can be brought to a
more convenient form. We set $\eta_S=0$. This implies that the
behavior of the system can be controlled by changing the temperature
gradient in the layer, while the salinity gradient remains constant
and equal to the critical value. The coefficient $\alpha_R$
($i\alpha_1/\eta=\alpha_R+i\alpha_I$) is eliminated from the
equation by changing the dependent variable by the formula
$A=A'\exp(-i\alpha_R{\eta}{T_2})$. Equation (\ref{CGLE}) then
becomes
\begin{eqnarray} \label{PNSE}
i\partial_{T_2} A' + \gamma_R\partial_X^2 A' - \beta_R A'|A'|^2
 = i\alpha_I\eta A' + i\gamma_I\partial_X^2 A'.
\end{eqnarray}
Here
\begin{eqnarray}
& & \alpha_R = \frac{3}{4}\pi^2\frac{\sigma+\tau}{1-\tau}\,
 \frac{\Omega^2+1}{\Omega^2+(1+\tau+\sigma)^2}
 \left( \Omega+ \frac{\tau (1+\tau+\sigma)}{\Omega} \right),
 \nonumber \\
& & \alpha_I = \frac{3}{4}\pi^2\frac{(\sigma+\tau)(\sigma+1)}
 {1-\tau}\,\frac{\Omega^2+1}{\Omega^2+(1+\tau+\sigma)^2},
 \nonumber \\
& & \beta_R = \frac{\pi^2}{8\Omega},  \qquad
\gamma_R = \Omega - 2\frac{(\sigma+\sigma\tau+\tau)\Omega^2 +
 \sigma\tau (1+\tau+\sigma)}{\Omega[\Omega^2+(1+\tau+\sigma)^2]},
 \nonumber \\
& & \gamma_I = 2\frac{(\sigma+\tau)(1+\tau+\sigma+\tau\sigma)}
  {\Omega^2+(1+\tau+\sigma)^2}. \nonumber
\end{eqnarray}
%
%
\begin{figure}[ptb]
\centering \includegraphics[width=0.7\textwidth,angle=-90]{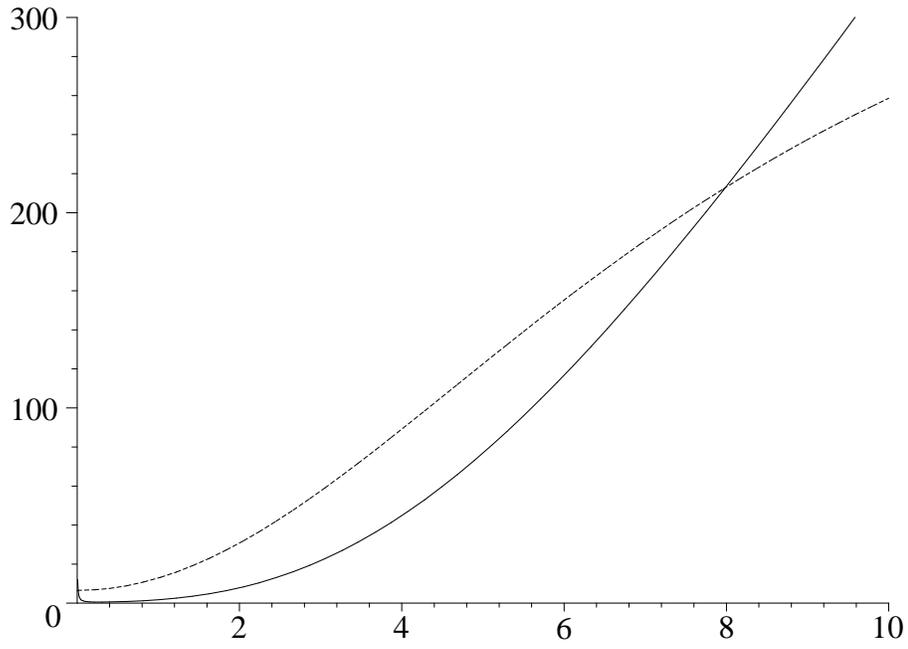}
\caption{Coefficients $\alpha_R(\Omega)$ (solid line) and
$\alpha_I(\Omega)$ (dashed line) in Eq.~(\ref{PNSE}), $\sigma=7$,
$\tau=1/81$.}\label{fig2ari}
\end{figure}
\begin{figure}[ptb]
\centering \includegraphics[width=0.7\textwidth,angle=-90]{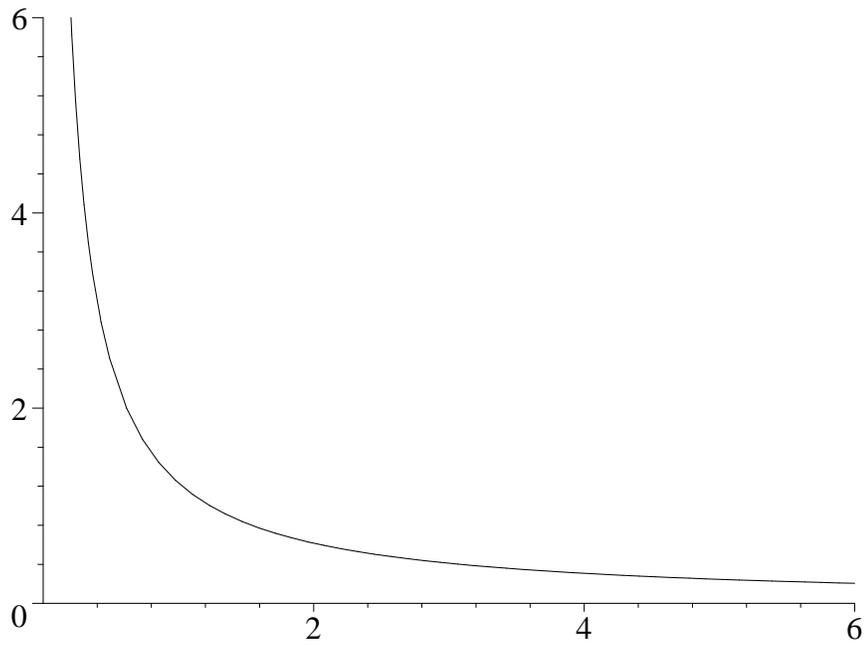}
\caption{Coefficient $\beta_R(\Omega)$ in Eq.~(\ref{PNSE}),
$\sigma=7$, $\tau=1/81$.}\label{fig2br}
\end{figure}
\begin{figure}[ptb]
\centering \includegraphics[width=0.7\textwidth,angle=-90]{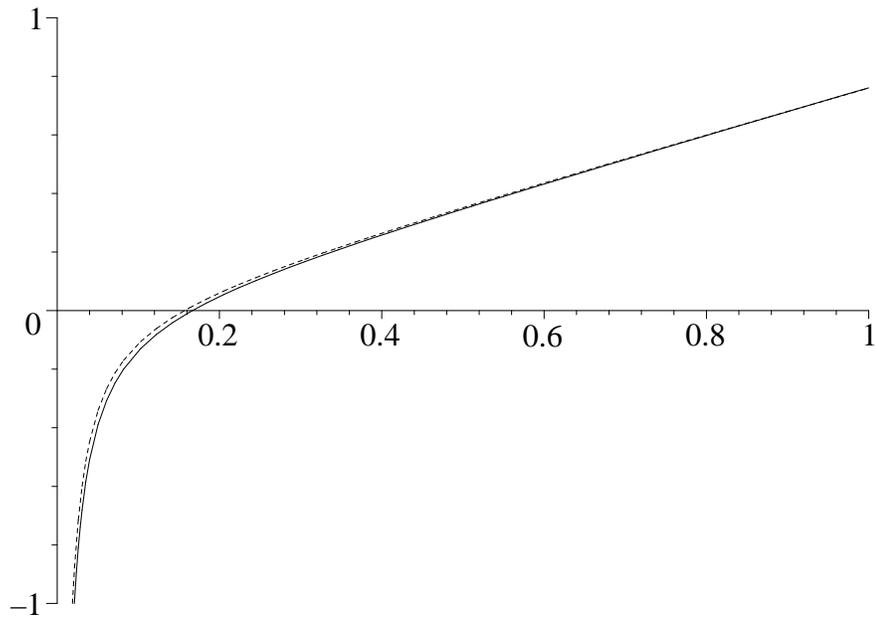}
\caption{Coefficient $\gamma_R(\Omega)$ in Eq.~(\ref{PNSE}),
$\sigma=7$, $\tau=1/81$. Dashed line is a two-term approximation for
a small $\Omega$. }\label{fig2gr}
\end{figure}
\begin{figure}[ptb]
\centering \includegraphics[width=0.7\textwidth,angle=-90]{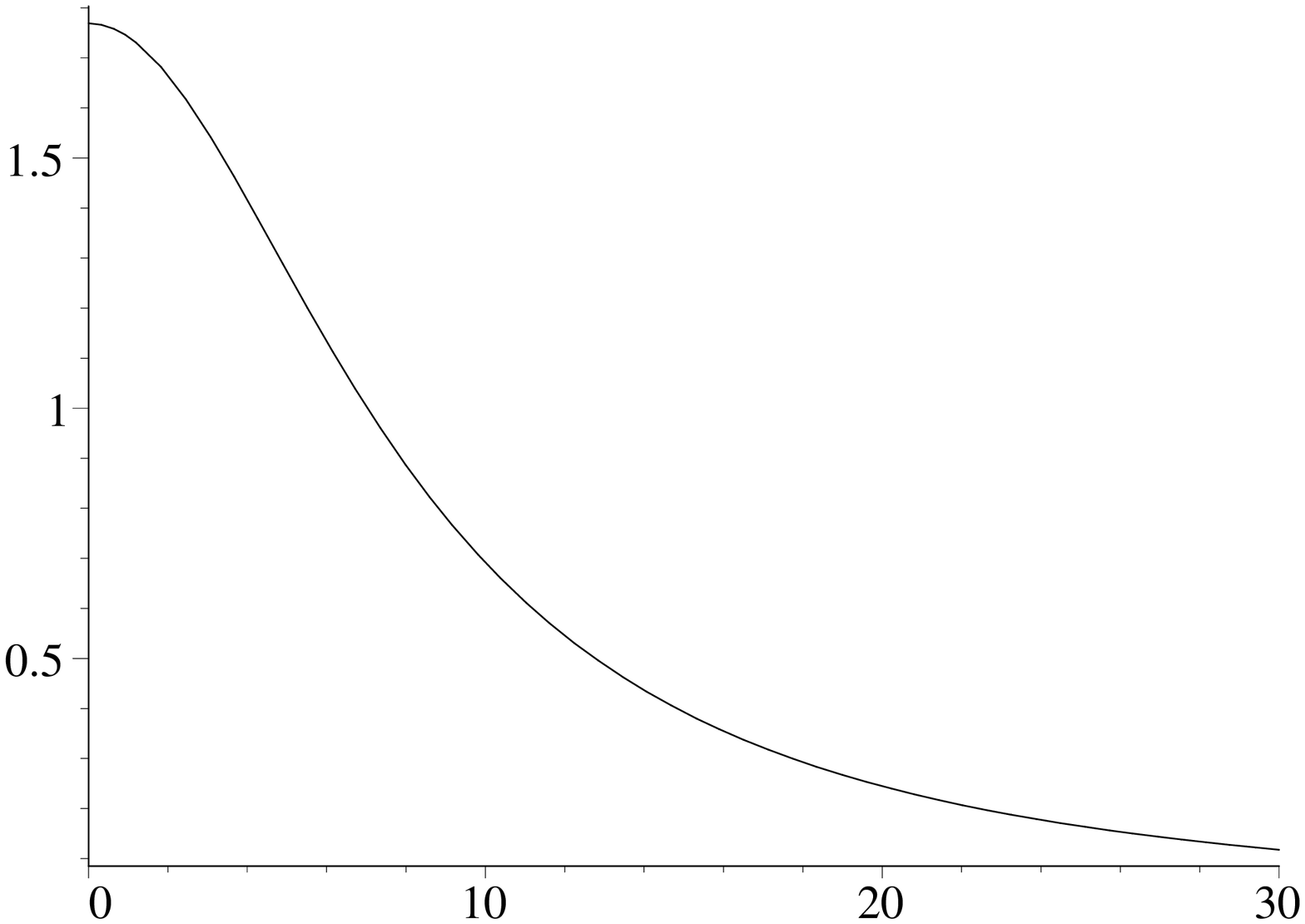}
\caption{Coefficient $\gamma_I(\Omega)$ in Eq.~(\ref{PNSE}),
$\sigma=7$, $\tau=1/81$.}\label{fig2gi}
\end{figure}

Thus, the amplitude equation becomes a nonlinear Schr\"odinger
equation (NSE) with perturbation on the right side. We note that the
coefficient values in this equation do not coincide with those given
by Bretherton and Spiegel \cite{Bretherton:1983}, who studied an
equation of the type (\ref{CGLE}) obtained by the method of
expansion of a linear dispersion relation in the neighborhood of a
critical wavenumber.

It is known \cite{Ablow:1987} that for a dissipatively perturbed
NSE, perturbation changes the form of solutions only slightly. As a
result, as the perturbing terms tend to zero, these solutions become
solutions of an NSE with no perturbed right side, which can be
solved in general form by the method of the inverse scattering
problem \cite{Dodd:1988}. If an NSE has soliton solutions (both
envelope solitons and solitons above a field of finite density), Eq.
(\ref{PNSE}) has solutions of the same form with rather small
perturbing terms. The type of NSE is determined by the sign of the
second derivative. In this case, the sign of the coefficient
$\gamma_R$ varies with change in $\Omega$ from zero to infinity.
Hence, in the problem considered, two types of NSE are possible:

\medskip

--- for $\Omega\rightarrow 0$
\begin{eqnarray}
& & \gamma_R = - \frac{2\sigma\tau}{1+\tau+\sigma}\,\Omega^{-1} +
    \nonumber \\
& & \qquad + \left(1-\frac{2(\tau+\sigma+\tau\sigma)}
 {(1+\tau+\sigma)^2} + \frac{2\tau\sigma}
 {(1+\tau+\sigma)^3}\right)\Omega+O(\Omega^3), \nonumber \\
& & \gamma_I = 2 + \frac{2}{1+\tau+\sigma}\left( \tau\sigma-1
- \frac{\tau\sigma}{1+\tau+\sigma} \right)+O(\Omega^2); \nonumber
\end{eqnarray}

--- for $\Omega\rightarrow \infty$
\begin{eqnarray}
& & \gamma_R = \Omega-2(\tau+\sigma+\tau\sigma)\Omega^{-1}
 + O(\Omega^{-3}), \nonumber \\
& & \gamma_I = 2(\tau+\sigma)(1+\tau+\sigma+\tau\sigma)
 \Omega^{-2} + O(\Omega^{-4}). \nonumber
\end{eqnarray}
In the limit $\Omega=0$, the coefficient $\gamma_R$ becomes infinity
and Eq. (\ref{PNSE}) loses meaning. This limiting case corresponds
to the double zero point of the dispersion relation. The amplitude
equation in the $\varepsilon^2$-neighborhood of this point will be
deduced below. As $\Omega$ increases from zero to infinity,
$\gamma_R$ changes sign, whereas $\gamma_I$ decreases monotonically,
remaining always positive. The frequency $\Omega_0$ at which
$\gamma_R$ vanishes is determined from the formula
\begin{eqnarray}
\Omega^2_0 = \frac{1}{2}(1+\sigma^2+\tau^2)
\left(\sqrt{1+\frac{8\sigma\tau(1+\tau+\sigma)}
{(1+\sigma^2+\tau^2)^2}} - 1 \right). \nonumber
\end{eqnarray}
\begin{figure}[ptb]
\centering \includegraphics[width=0.7\textwidth,angle=-90]{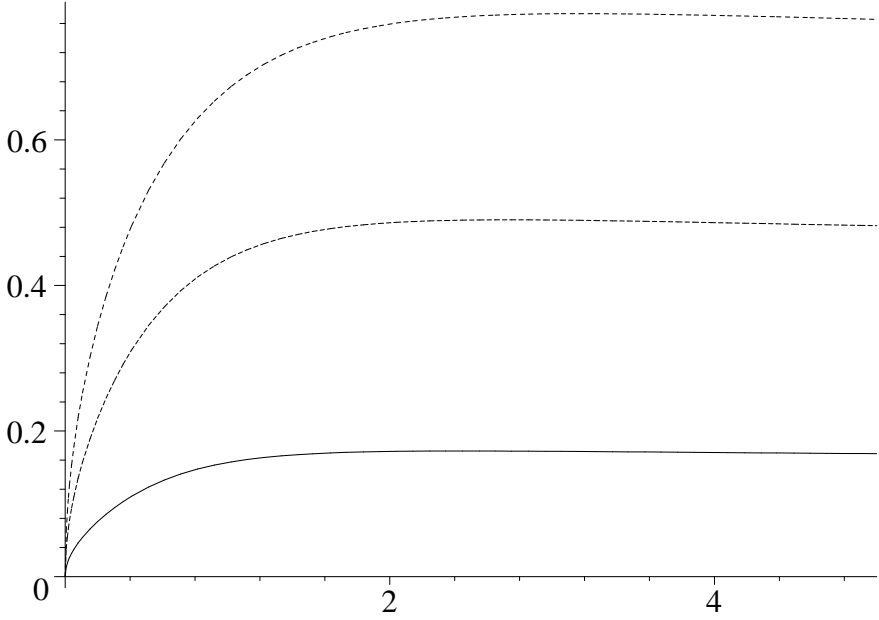}
\caption{Frequency $\Omega_0(\tau,\sigma)$ at which $\gamma_R$
vanishes, for $\tau=1/81$ (solid line), $\tau=1/10$ (dashed line),
$\tau=1/4$ (dotted line) and $\sigma$ in the range $0-5$.}\label{w0}
\end{figure}
For rather large $\sigma$ or small $\tau$, this formula has the
asymptotic form $\Omega^2_0 \approx
 2\tau\sigma(1+\tau+\sigma)/(1+\tau^2+\sigma^2)$. In the case, when
$\sigma=7$, $\tau=1/81.$ we have $\Omega_0\approx 0.1663778362$. For
the other values of $\sigma$ and $\tau$ see Fig.~\ref{w0}. In the
limit of the infinite $\sigma$ it is true $\Omega_0 = \sqrt{2\tau}$.


\section{Transformation to a Nonlinear Schr\"odinger Equation.}
\begin{figure}[tbh]
\centering \includegraphics[width=0.7\textwidth,angle=-90]{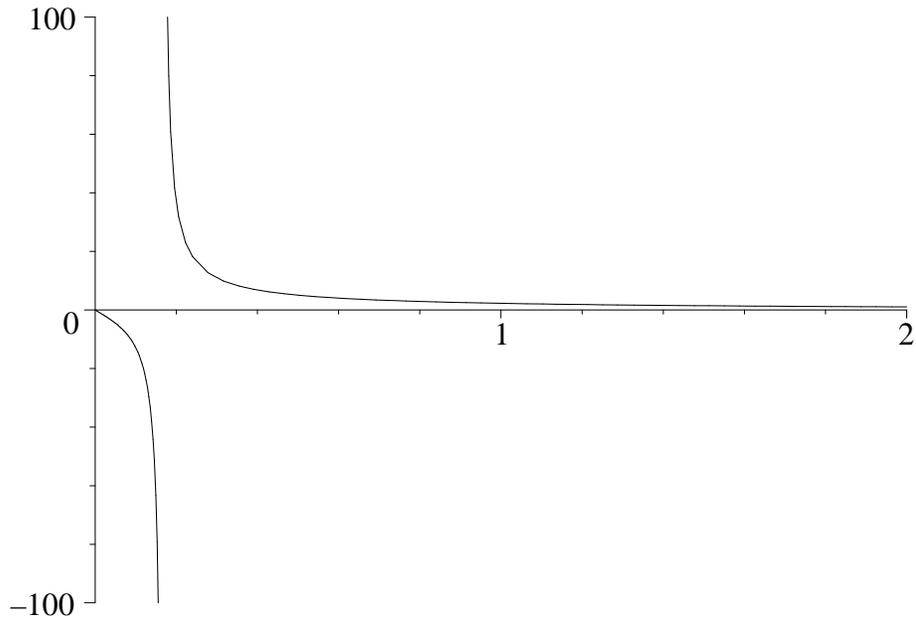}
\caption{Coefficient $\mu(\Omega)$ in Eq.~(\ref{eqs}), $\sigma=7$,
$\tau=1/81$.}\label{fig3mu}
\end{figure}

We consider two cases where the amplitude equation derived above
becomes an NSE. Using the substitution
\begin{eqnarray}
A=\sqrt{{\alpha_I}/{\beta_R}}\exp[-i(\alpha_R+\alpha_I\rho^2)
T_2] F(\alpha_I{T_2},\sqrt{\alpha_I/\gamma_R}X), \nonumber
\end{eqnarray}
where $\rho$ is a positive constant, we bring Eq. (\ref{PNSE}) to
the form
\begin{eqnarray}\label{eqs}
iF_T+F_{XX}-F(|F|^2-\rho^2)=i{\eta}F+i\mu{F_{XX}},
\end{eqnarray}
where $\mu=\gamma_I/\gamma_R$ (Fig.~\ref{fig3mu}). Here and below,
the subscripts $T$ and $X$ denote partial derivatives with respect
to the slow time $T_2$ and the $X$ coordinate, respectively. The
coefficient $\mu$ tends to zero with increase in $\Omega$ according
to the asymptotic relation
\begin{figure}[tbh]
\centering \includegraphics[width=0.7\textwidth,angle=-90]{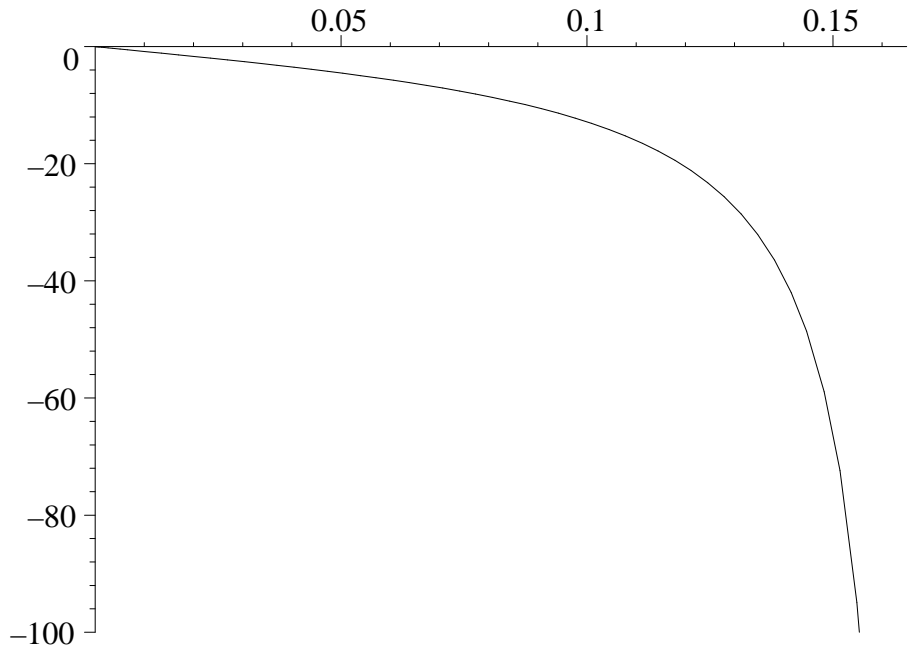}
\caption{Coefficient $\mu(\Omega)$ in Eq.~(\ref{eqs}) for a small
$\Omega$, $\sigma=7$, $\tau=1/81$.}\label{fig3mul}
\end{figure}
$\mu \approx 2(\tau+\sigma)(1+\tau+\sigma+\tau\sigma)
{\Omega^{-3}}$.
In addition, in the immediate vicinity of the Hopf bifurcation point
(in the $\varepsilon^3$ neighborhood), the first term on the right
side of Eq. ($\ref {eqs}$) can be eliminated. The second term can
also be ignored if the frequency $\Omega$ is sufficiently high. As a
result, Eq. (\ref{PNSE}) becomes the NSE
\begin{eqnarray}
iF_T+F_{XX}-F(|F|^2-\rho^2)=0. \nonumber
\end{eqnarray}
This equation has solutions that are known as solitons of finite
density or ``dark" solitons \cite{Tahtaj:1986}:
\begin{eqnarray} \label{DARK}
& F = \displaystyle \rho\,\frac{\exp(i\zeta)+\exp\Phi}
  {1+\exp\Phi}, \quad |F|^2 = \displaystyle
  \rho^2\left(1-\frac{\sin^2(\zeta/2)}{\ch^2\Phi}\right), & \\
& \Phi = \displaystyle - \rho T \sin\zeta
  \pm(X-X_0)\sqrt{2\rho}\sin(\zeta/2). & \nonumber
\end{eqnarray}
The parameters $\zeta$ and $X_0$ characterize the width and initial
position of the soliton, respectively.

Thus, the present investigation shows that for the physical system
considered, along with other solutions, there can be solutions of
the type of ``dark" solitons, and this is true in the limit of high
Hopf frequencies. Apparently, double-diffusive convection at high
Hopf frequencies can occur in ocean systems. An example of these
systems is a so-called {\it thermohaline staircase}
\cite{Marmorino:1990}. Inversions of a thermohaline staircase often
have stratification parameters, which correspond to the beginning of
diffusive convection, and the Hopf frequency $\Omega$ is of the
order of $10^3$--$10^5$.

When the Hopf frequency tends to zero, Eq. (\ref{PNSE}) takes a
different asymptotic form. In this case, we set
\begin{eqnarray}
A = \sqrt{\alpha_I/\beta_R}\exp(-i\alpha_R T_2)
F(\alpha_I{T_2},-\sqrt{\alpha_I/\gamma_R}X). \nonumber
\end{eqnarray}
Then,
\begin{eqnarray}
iF_T-F_{XX}-F|F|^2=i\eta F+i\mu{F_{XX}}, \nonumber
\end{eqnarray}
where $\mu$ has the following low-frequency asymptotic form:
\begin{eqnarray}
\mu \approx - \Omega\left(1+\frac{\tau+\sigma}{\tau\sigma} -
\frac{1}{1+\tau+\sigma}\right). \nonumber
\end{eqnarray}
Thus, $\mu\rightarrow 0$ as $\Omega\rightarrow 0$
(Fig.~\ref{fig3mul}). As in the previous case, the first term on the
right side of the equation can be eliminated by assuming that the
system is in the immediate vicinity (in the $\varepsilon^3$
neighborhood) of the Hopf bifurcation point. Then, again, Eq.
(\ref{PNSE}) takes the form of an NSE:
\begin{eqnarray}
iF_T=F_{XX}+F|F|^2.  \nonumber
\end{eqnarray}
This equation has well-known solutions in the form of envelope
solitons.

It is interesting that localized wave packets, with which soliton
solution can be compared, were observed in experiments on convection
of binary mixtures at rather low Hopf frequencies (see, for example,
\cite{Predtechensky:1994a,Kolodner:1992}).


\section{Equations at the Taylor Bifurcation Points and Double Zero Point.}
We consider the case of Taylor bifurcation or bifurcation to steady
roll-type convection. On the straight line on which this bifurcation
occurs, the dispersion relation has a first-order root. For terms of
the order of $O(\varepsilon^2)$, the equation has the form
$\partial_{T_1}A=0$, i.e., the amplitude does not depend on the slow
variable $T_1$. For terms of the order of $O(\varepsilon^3)$ of the
functions ${\bf Q}_3$, we obtain the expressions
\begin{eqnarray}
& & Q_{A3}=
\frac{9}{4}\sigma\pi^4 [r^{*}_{T}(\eta_S-\eta)-\eta_S]A
+\frac{3\pi}{\sqrt{2}}\partial_{T_2}A-
\frac{7\sigma\pi^2}{2}\partial^2_{X_1}A, \nonumber  \\
& & Q_{T3}=\frac{i\pi}{6\sqrt{2}}\left(A|A|^2+
\frac{4}{\pi^2}\partial_{T_2}A-
\frac{20}{3\pi^2}\partial^2_{X_1}A\right), \nonumber \\
& & Q_{S3}=\frac{i\pi}{6\tau^2\sqrt{2}}\left(A|A|^2+
\frac{4\tau}{\pi^2}\partial_{T_2}A-
\frac{20\tau^2}{3\pi^2}\partial^2_{X_1}A\right).  \nonumber
\end{eqnarray}
Substituting these formulas into the compatibility condition, we
have the amplitude equation
\begin{eqnarray} \label{NWE}
\partial_{T_2}A=\alpha_3 A - \beta_3 A|A|^2
- \gamma_3\partial^2_{X_1}A,
\end{eqnarray}
where
\begin{eqnarray}
& \alpha_3 = \displaystyle \frac{3}{2}\pi^2\tau
  \frac{r^{*}_{T}(\eta_S-\eta)-\eta_S}
  {r^{*}_{T}(1-\tau)-(1+\tau/\sigma)}, \qquad
\beta_3 = \displaystyle  \frac{\pi^2}{4\tau}
  \frac{r^{*}_{T}(1-\tau^2)-1}{r^{*}_{T}(1-\tau)
  - (1+\tau/\sigma)},& \nonumber \\
& \gamma_3 = \displaystyle \frac{4\tau}
  {r^{*}_{T}(1-\tau)-(1+\tau/\sigma)}. & \nonumber
\end{eqnarray}
This equation is similar in form to the equation derived in
\cite{Newell:1969} and reduces to it if a salinity gradient is
absent.

We consider the $\varepsilon^2$ neighborhood of the point of
intersection of the straight lines on which Hopf and Taylor
bifurcations are observed. At this point, the dispersion relation
has a second-order root (Takens-Bogdanov bifurcation). As noted
above, for the case of the most unstable convective mode, the
equation obtained for terms of the order of $O(\varepsilon^2)$ is
satisfied identically. Therefore, it is not necessary to use the
variable TI or to introduce other slow variables. For terms of the
order of $O(\varepsilon^3)$ of the functions ${\bf Q}_3$, we obtain
the expressions
\begin{eqnarray}
& & Q_{A3} =
    \frac{9\pi^4}{4(1-\tau)}
    [(\sigma+1)\eta_S - (1+{\sigma}/{\tau})\eta] A
    - \frac{i\pi}{\sqrt{2}}\partial_{X_1}\partial_{T_1} A
    - \frac{7\sigma\pi^2}{2}\partial^2_{X_1} A, \nonumber \\
& & Q_{T3} = \frac{i\pi}{6\sqrt{2}}\left(A|A|^2
    - \frac{8}{3\pi^4}\partial^2_{T_1} A
    - \frac{20}{3\pi^2}\partial^2_{X_1}A\right)
    + \frac{2}{9\pi^2}\partial_{X_1}\partial_{T_1}A, \nonumber \\
& & Q_{S3} = \frac{i\pi}{6\tau^2\sqrt{2}}\left(A|A|^2
    - \frac{8}{3\pi^4}\partial^2_{T_1} A
    - \frac{20\tau}{3\pi^2}\partial^2_{X_1}A\right)
    + \frac{2}{9\pi^2\tau}\partial_{X_1}\partial_{T_1}A.\nonumber
\end{eqnarray}
After substitution of these expressions into the condition of the
absence of secular terms, we obtain the equation
\begin{eqnarray} \label{FI4}
\partial^2_{T_1}A-c^2\partial^2_{X_1}A=\alpha_2 A
+ \beta_2 A|A|^2,
\end{eqnarray}
where
\begin{eqnarray}
c^2 = \frac{6\pi^2\sigma\tau}{1+\tau+\sigma}, \qquad
   \beta_2 = \frac{3}{8}\pi^4, \qquad
\alpha_2 = \frac{9}{4}\pi^4\tau^2\frac{(1+\sigma/\tau)\eta
 - (1+\sigma)\eta_S}{(1-\tau)(1+\tau+\sigma)}. \nonumber
\end{eqnarray}
Equations of this type are known as $\varphi^4$-equations, and they
cannot be integrated accurately by the method of the inverse
scattering problem \cite{Dodd:1988}. Some
papers~\cite{Brand:1984,Cross:1988a,Knobloch:1986a} consider
amplitude equations at the double point for the convection of binary
mixtures. According to \cite{Knobloch:1980}, the results obtained
for thermohaline convection are extended to the case of convection
of binary mixtures, where it is necessary to allow for the
thermodiffusion effect. Therefore, for the last case, all the
equations at bifurcation points derived in the present paper are
valid with the parameters of the problem converted accordingly
(Prandtl, Lewis, and Rayleigh numbers). Knobloch
\cite{Knobloch:1986a} obtained an amplitude equation at the double
zero point that has the form $\partial^2_{T_1}A=C_1 A + C_2 A|A|^2$
in the main order in $\varepsilon$ ($C_1$ and $C_2$ are constants).
Equation (\ref{FI4}) can be regarded as its extension to the case of
spatial modulations. Brand et al. \cite{Brand:1984} gives another
amplitude equation at the double zero point, which includes a term
with a third derivative of the form $\partial_t\partial^2_x A$.
Therefore, it differs from the equations derived by the
multiple-scale expansion method used in the present paper.


\section{Conclusions.}
1. The derivative expansion method is used to derive amplitude
equations for a system with thermohaline convection in the
neighborhood of the main bifurcation points characteristic of this
system. In particular, within the framework of a unified approach,
we obtained the complex Ginzburg-Landau equation (\ref{CGLE}) in the
case of Hopf bifurcation, the Newell-Whitehead equation (\ref{NWE})
in the case of Taylor bifurcation, and Eq. (\ref{FI4}) of the
$\varphi^4$ type in the neighborhood of the double zero point of the
dispersion relation.

2. Analytic expressions for the coefficients of the equations
considered are given. For the equation in the neighborhood of the
Hopf bifurcation points, the formulas specifying its coefficients
refine the previous results of \cite{Bretherton:1983}. For the other
two equations, such formulas, as far as we know, have not been
previously reported in the literature.

3. It is shown that, for low and high frequencies, the amplitude
equation in the neighborhood of the Hopf bifurcation points reduces
to the perturbed nonlinear Schr\"odinger equations (\ref{PNSE}) with
characteristic solutions in the form of envelope solitons. In the
high-frequency limit, the type of ``dark" solitons (\ref{DARK}) are
characteristic of the examined physical system.

4. The equation of the type of $\varphi^4$ derived at the double
zero point of the dispersion relation can be regarded as an
extension of the equation obtained in \cite{Knobloch:1986a} to the
case of slow spatial modulations of the amplitude.


\end{document}